\documentstyle[12pt]{article}
\newcommand{\bsigma}{\mbox{\boldmath $\sigma$}}

\newcommand{\beq}{\begin{equation}}
\newcommand{\eeq}{\end{equation}}
\newcommand{\bea}{\begin{eqnarray}}
\newcommand{\eea}{\end{eqnarray}}

\begin{document}
\thispagestyle{empty}
\vspace*{-15mm}
\baselineskip 10pt
\begin{flushright}
\begin{tabular}{l}
{\bf OCHA-PP-175}\\
{\bf April 2001}\\
{\bf hep-th/0104241}
\end{tabular}
\end{flushright}
\baselineskip 24pt 
\vglue 10mm 

\begin{center}
{\LARGE\bf 
4d Gauge Theory and Gravity Generated from 3d Ones \\
at High Energy
}
\vspace{7mm}

\baselineskip 18pt 
{\bf
Akio SUGAMOTO}
\vspace{2mm}

{\it 
 Department of Physics, Ochanomizu University, \\
 2-1-1, Otsuka, Bunkyo-ku, Tokyo 112-8610, Japan
}\\
\vspace{10mm}
\end{center}
\begin{center}
{\bf Abstract}\\[7mm]
\begin{minipage}{14cm}
\baselineskip 16pt
\noindent
Dynamical generation of 4d gauge theories and gravity at low 
energy from the 3d ones at high energy is studied, based on 
the fermion condensation mechanism recently proposed by 
Arkani-Hamed, Cohen and Georgi. For gravity, 4d Einstein gravity is
generated from the multiple copy of the 3d ones, by referring to the two
form gravity.  Since the 3d Einstein action without matter coupling is
topological, ultraviolet divergences are less singular in our model.  In
the gauge model, 
matter fermions are introduced on the discrete lattice following Wilson. 
Then, the 4d gauge interactions are correctly generated from 
the 3d theories even in the left-right asymmetric theories of 
the standard model. In order for this to occur, the Higgs fields 
as well as the gauge fields of the extra dimension should be 
generated by the fermion condensates.  Therefore, the generation 
of the 4d standard model from the multiple copy of the 3d ones 
is quite promising. To solve the doubling problem in the weak interaction
sector, two kinds of discrete lattices have to be introduced separately for
L- and R-handed sectors, and the two types of Higgs fields should be
generated. 
\end{minipage}
\end{center}

\newpage
\baselineskip 18pt 
\def\thefootnote{\fnsymbol{footnote}}
\setcounter{footnote}{0}

\section{Introduction}
In a recent paper,  Arkani-Hamed, Cohen and Georgi~\cite{Georgi} 
have proposed an interesting model in which 5d gauge theory is 
generated dynamically from the 4d asymptotically free gauge 
theory at high energy.  
It is by no means new that the extra gauge potential is generated 
dynamically at low energy as the  fermion pair condensates, but 
the concept that the dynamical generation of the extra (5th) 
dimension at the same time is completely new.

We will firstly reformulate their model by generating explicitly 
the D+1 dimensional gauge theory action from the D dimensional one.  
In the course of the study, generation of the space-like dimension 
is found to be easier than that of the time-like one.

Next, we propose a model in which 4d gravity is generated at low 
energy by the fermion pair condensates from the 3d ones at high 
energy.  
By referring to the 2-form gravity~\cite{2-form gravity},
\cite{condensation of string}, 4d Einstien gravity is shown to be generated

from the multiplication of the 3d Einstein gravities. 
As is well known 3d Einstein gravity without matter coupling has no
dynamical degrees of 
freedom (being a topological theory~\cite{Witten}), so that the ultraviolet
divergences in our model are less singular.

Thirdly, we introduce matter fermions into the 3d gauge theory 
and see how 4d matter couplings are induced.  
We use the Wilson's method~\cite{lattice theory} of introducing 
fermion on the discrete lattice and find that it works naturally 
well in our generation of 4d theory from 3d ones. 
Especially, the action added by Wilson can be reproduced, if 
the fermion condensates corresponding to the Higgs scalars are 
generated.  
Although the doubling problem~\cite{doubling problem} does not 
occur in the left-right symmetric theories such as QED and QCD, 
it occurs in the left-right asymmetric QFD.
Nevertheless the possibility of generating the 4d standard model 
from the simple multiplication of the 3d ones at high energy is 
quite promising. 
To solve the doubling problem, two kinds of discrete lattices have to be
introduced separately for L- and R-handed fields, and the two types of
Higgs fields should be generated.
\section{
Dynamical generation of  (D+1)d gauge theory from Dd one 
at high energy}
Following the model by Arkani-Hamed et al.~\cite{Georgi}, we begin 
with the gauge theory having gauge groups $(G)^{N} \times (G_{s})^{N}$, 
the direct product of independent $N$ sets of  group $G \times G_{s}$.  
For simplicity we can restrict $G$ to $SU(m)$ and $G_{s}$ to $SU(n_{s})$.
These groups are ordered linearly with the index 
$\{1, \frac{3}{2}, 2, \frac{5}{2}, \cdots, n, n+\frac{1}{2}, n+1, 
\cdots, N-\frac{1}{2}, N\}$. 
These indices become an extra discrete coordinate later. 
At integer index $n$ the group $G(n)$ and its gauge fields $A(n)$ 
are assigned, while at half integer index $n+\frac{1}{2}$ 
the strong group $G_{s}(n+\frac{1}{2})$ and its gauge fields 
$A(n+\frac{1}{2})$ are assigned.  
There are Weyl fermions $\chi(n, n+\frac{1}{2})$ and 
$\chi(n+\frac{1}{2}, n+1)$ which connect, respectively, 
two groups $(G(n), G_{s}(n+\frac{1}{2}))$, and 
$(G_{s}(n+\frac{1}{2}), G(n+1))$. 
Namely, the representation of $\chi(n, n+\frac{1}{2})$ is 
non-trivial $( \{m \}, \{\bar{n}_{s}\})$ only under the adjacent set 
of groups $(G(n), G_{s}(n+\frac{1}{2}))$, while that of 
$\chi(n+\frac{1}{2}, n+1)$ is non-trivial $(\{n_{s}\}, \{\bar{m}\} )$ 
only under $(G_{s}(n+\frac{1}{2}), G(n+1))$.  
If at the energy lower than $\Lambda_{s}$, the coupling constants 
$g_{s}(n)$ of the groups $G_{s}(n)$ at the position 
$n+\frac{1}{2} (n=0-(N-1))$ become commonly strong, then the fermion 
pair condensates 
$\langle \chi(n,n+\frac{1}{2}) \chi(n+\frac{1}{2}, n+1)\rangle$ 
are formed at the lower energy than a common energy scale 
$\Lambda_{s}$.  
This condensate is a singlet under $G_{s}(n+\frac{1}{2})$, 
but transforms as $(\{m\}, \{\bar{m}\})$ under the groups 
$(G(n), G(n+1))$. 
Therefore, after the condensates are formed, the independently 
prepared groups $G(n) (n=1, \cdots, N)$ with labels $n=1, \cdots, N$ 
(or at different points $n=1, \cdots, N$) are unified by the gauge 
principle.

To clarify this point furthermore let us denote the condensate as
\beq
\frac{1}{2\pi (f_{s})^{D-1}} 
\langle \chi(n, n+\frac{1}{2})\chi(n+\frac{1}{2}, n+1)\rangle 
= U(x; n, n+1) = e^{iaA_{0}(x, n)}, 
\label{condensation}
\eeq
where $U(x; n, n+1)$ is a $m \times m$ unitary matrix and $A_{0}(x, n)$ 
is a $m \times m$ hermitian matrix playing the role of gauge potential
(connection) in the direction of the extra dimension.  
Here, we adopt the convention in which the index $\{0 \}$ represents 
the extra dimension, while $i=1, \cdots, D$ denote the original 
$D$ dimensions. 

Now in the low energy effective theory the following gauge invariant 
terms can be generated:
\beq
\Delta S = (f_{s})^{D-2}~\int d^{D}x ~\sum^{N}_{n=1} tr
\left[\left(D_{i}U(x; n, n+1)\right)^{\dagger}\left(D^{i}U(x; n,
n+1)\right)\right],
\eeq
where $D_{i}$ is the covariant derivative of the relevant gauge groups
$G(n) \times G(n+1)$, 
\beq
D_{i}U=\partial_{i}U +iA_{i}(x, n)U(x; n, n+1) -iU(x; n, n+1)A_{i}(x, n+1).
\eeq
Using Eq.(\ref{condensation}) and the expansion in $a$, we have
\bea
& &D_{i}U(x; n, n+1)~=~ia \biggl(
\partial_{i}A_{0}(x, n)-\frac{1}{a} 
    \{ A_{i}(x, n+1) -A_{i}(x, n) \}  \nonumber \\ 
& &
+ i[A_{i}(x, n)A_{0}(x, n)-A_{0}(x, n)A_{i}(x, n+1)] \biggr)
+ \cdots, 
\label{extra gauge field}
\eea
which can be identified to the field strength of $\{i, 0\}$ component, 
$iaF_{i0}+\cdots$.  
Now the newly generated action becomes
\beq
\Delta S = f_{s}^{D-2} \int d^{D}x \sum^{N}_{n=1} a^{2}
(F_{i0})^{\dagger}(F^{i}_{~0}) +\cdots.
\eeq
The generated action is combined with the original action
\beq
S=-\frac{1}{2 (g_{D})^{2}}  \int d^{D}x \sum_{n=1}^{N} tr \left( F_{ij}(x,
n) F^{ij}(x, n) \right), 
\label{3D gauge action}
\eeq
leading to the $D+1$ dimensional gauge theory of the group $G$:
\bea
S+\Delta S &=& 
-\frac{1}{2(g_{D+1})^2} 
\int d^D x ~~a\sum_{n=1}^{N}~~tr(F_{MN}(x, n) F^{MN}(x, n)) 
\nonumber \\
&&~~ + \mbox{(higher order terms in $a$)}, 
\label{D+1 gauge theory}
\eea
where $M$ and $N$ run from $0, 1, \cdots, D$.

For the result (\ref{D+1 gauge theory}) to occur, 
the extra dimension should be 
"space-like", and the following relations should hold: 
\beq
a=\frac{1}{g_{D}(f_{s})^{(D-2)/2}} ~~~\mbox{and}
~~~(g_{D+1})^{2}=a(g_{D})^{2}. 
\label{conditons}
\eeq
In order to generate the "time-like" dimension, the sign of 
the generated action $\Delta S$ should be altered.  
This is usually impossible, since the collective excitations 
such as $U(x; n, n+1)$ increase usually the energy of the system.  
It may, however, possible if we can consider the scalar field 
$U(x; n, n+1)$ to be dilatonic ones, since the kinetic terms of 
the dilatonic fields can have the opposite sign compared to those 
of the usual scalar fields. 

Now the propagator of the D+1 dimensional gauge boson in the Feynman 
gauge reads 
\beq
D_{MN} (p, p_{0})=\frac{-i
g_{MN}}{\sum_{i=1}^{D}p_{i}p^{i}-(\frac{2}{a})^{2}\sin^{2} 
(\frac{ap_{0}}{2})} 
+\mbox{(higher order terms in $a$)},
\eeq
where $p_{0}$ is the momentum of the extra discrete space generated,
satisfying $-\frac{\pi}{a}< p_{0}<\frac{\pi}{a}$.  
If $a \rightarrow 0$ and $N \rightarrow \infty$, then the continuous 
$D+1$ dimensional gauge theory is obtained, but at finite $a$, 
modification from the continuous 4d gauge theory will manifest 
itself at higher energy around $1/a$ or $g_{D}(f_{s})^{\frac{D-1}{2}}$.
\section{
Dynamical generation of 4d gravity from the 3d one at high energy
}
In this section we apply the mechanism of the last section to gravity. 
There are various ways to formulate Einstein gravity. 
Among them we refer to the so called 2-form gravity or the Ashtekar
formalism of it in 4d~\cite{2-form gravity}, since 3d and 4d gravities 
are treated similarly in this formalism, based on the gauge principle. 

At high energy we start with the sum of $N$ sets of independent 
3d gravity actions based on the local Lorentz group $(G_{L} )^{N}$.  
Here, an important point is that the gauge group $(G_{L} )$ is 
taken to be 4d local Lorentz group SO(1,3) (or SO(4)) and not 
to be 3d SO(1, 2) (or SO(3)). 
To simplify the following discussion, we use 
$SO(4)=SU(2)\times\overline{SU(2)}$ as this gauge group $G_{L}$.  
Then, we have the action at high energy as
\begin{equation}
S_{G} =\frac{1}{2 (\kappa_{3})^{2}}\sum_{n=1}^{N}\int d^3x
\frac{1}{2}\epsilon^{ijk} B^{AB}_{i}(x, n)R^{AB}_{jk}(x, n).
\label{original 3D gravity action}
\end{equation}
Here, $B^{AB}_{i}(x, n)$ is the  $SO(4)$ gauge fields and 
$R^{AB}_{jk}$ is the $SO(4)$ field strength (Riemann curvature) of 
the spin connections $\omega^{AB}_{i}$ (gauge fields) defined by 
\begin{equation}
R^{AB}_{jk} = 
\partial_{[j}\omega^{AB}_{k]}+\omega^{AC}_{[j}\omega^{CB}_{k]}.
\end{equation}
The 3d gravity coupling $\kappa_{3}$ is related to the 3d Newton constant
$G_{3}$ as $(\kappa_{3})^{2}=8 \pi G_{3}$.

Now we will show that the action (\ref{original 3D gravity action}) 
is nothing but the $2N$ times multiple of the 3d Einstein gravity.  
To do this, we have to decompose $SO(4)$  adjoint field $F_{AB}$ 
into the adjoint fields $F^{a}$ and $\bar{F}^{a}$ of $SU(2)$ and 
$\overline{SU(2)}$, respectively: By using 't Hooft 
symbols~\cite{'t Hooft} $\eta^{a}_{AB}$ and $\bar{\eta}^{a}_{AB}$, 
or explicitly we have 
\bea
T_{AB} &=& \frac{1}{2} \left(\eta^{a}_{AB}T^{a} 
+ \bar{\eta}^{a}_{AB}\bar{T}^{a} \right),
~~~ \mbox{or explicitly} ~ \nonumber \\
T_{ab}&=&\frac{1}{2} \epsilon_{abc}(T^{a}+\bar{T}^{a}),
T_{0a}=\frac{1}{2} (-T^{a}+\bar{T}^{a}).
\eea
Our convention is that $A, B, \cdots$ run within 4d indices, 
while $a, b, \cdots$ run within 3d indices.

Then, we have the following decomposition: 
\begin{equation}
S_{G} =\frac{1}{2(\kappa_{3})^{2}} \sum_{n=1}^{N} 
\int d^3x \frac{1}{2} \epsilon^{ijk} 
\left(e^{a}_{i}(n)R^{a}_{jk}(n)
+\bar{e}^{a}_{i}(n)\bar{R}^{a}_{jk}(n) \right), 
\label{3D gravity}
\end{equation}
where the 3d vierbeins  $e^{a}_{i}(n)$ and $\bar{e}^{a}_{i}(n)$ 
are, respectively, $SU(2)$ and $\overline{SU(2)}$ decomposition 
of $B^{AB}_{i}(n)$.

Two terms in the action without and with "bar", correspond to 
$SU(2)$ and $\overline{SU(2)}$, respectively.  
They are at present independently duplicated in 3d gravity, but will be 
self-dual (chiral) and anti self-dual (anti chiral) parts of 
4d gravity.

So far the vierbeins $e^{a}_{i}$ and $\bar{e}^{a}_{i}$ and the 
spin connections $\omega_{i}^{a}$ and $\bar{\omega}_{i}^{a}$ are 
independent variables.  
After solving the equation of motion with respect to the spin 
connections (or eliminating the spin connections by path 
integration), we have the torsionless conditions:
\begin{equation}
D_i e_j^a - D_j e_i^a =
\bar{D}_i \bar{e}_j^a - \bar{D}_j \bar{e}_i^a =0, 
\end{equation}
which can be solved, giving the ordinary expression of 
$\omega_{i}^{a}$ in terms of $e^{a}_{i}$.  

Now we arrive at the $2N$ times duplication of 3d Einstein actions:
\begin{equation}
S_{G}=\frac{1}{2(\kappa_{3})^{2}}\sum_{n=1}^{N}\int d^3x
\left(e(n)R(n)+\bar{e}(n)\bar{R}(n)\right).
\end{equation}
As is shown by Witten~\cite{Witten}, this kind of theory 
can be rewritten to the Chern-Simons action based on the 
inhomogeneous Lorentz group ISO(3), which is the so called 
topological field theories without quantum degrees of freedom.

Coming back to the starting gravity action (\ref{original 3D 
gravity action}) at high energy, we will add the 2 component 
fermions $\lambda(n, n+\frac{1}{2})$ and 
$\lambda(n+\frac{1}{2}, n+1)$ transforming as 
$(\{4\}, \{\bar{n}_{s}'\})$ and $(\{n_{s}'\}, \{4\})$ 
under the groups $(G_{L},G_{s}')$ and $(G_{s}', G_{L})$, 
respectively.  
The strong group $G_{s}'$ is not necessarily the same one $G_{s}$ 
chosen to generate the 4d gauge theories.

In the same way as in the previous section, we may have the 
fermion pair condensates at energy lower than $\Lambda_{s}'$ by 
the strong interaction based on the group $G_{s}'(n)$. 
The condensates transform vector like under the groups 
$G_{L}(n) \times G_{L}(n+1)$, and are expressed by $N \times N$ 
orthogonal matrix, namely,
\begin{equation}
\frac{1}{2\pi f_{g}^{2}} \langle 
\lambda(n,n+\frac12)\lambda(n+\frac12, n+1) \rangle 
= O(x;n, n+1)=e^{a\omega_{0}(x, n)}.
\end{equation}
Then, similarly as before, we have the following expression,
\bea
& &D_i O(x, n, n+1)^{AB}  \nonumber \\ 
& & =a \biggl( \partial_i \omega^{AB}_0(x, n) - \frac{1}{a} 
\{\omega^{AB}_i(x, n+1) -\omega^{AB}_i(x, n)\} \nonumber \\ 
& & + \left[ \omega^{AC}_i(x, n) \omega^{CB}_0(x, n) 
- \omega_0^{AC}(x,n) \omega^{CB}_i(x, n+1) \right] 
+ \cdots \biggr)  \nonumber  \\ 
& & =a R^{AB}_{i0}(x, n) + \cdots. 
\eea
To approach the 4d gravity, introduction of the 4d vierbein is 
necessary, but so far we have only two sets of 3d vierbeins 
$e^{a}_{i}(n)$ and $\bar{e}^{a}_{i}(n)$ at each point $n$, the 
degrees of freedom of which are 18 at each point.  
Under a certain condition, $B^{AB}_{i}$ can be expressed in terms 
of the 16 4d vierbeins $E^{A}_{\mu}$ at each point 
as follows:
\begin{equation}
B^{AB}_{i}=B^{AB}_{0i}=\frac{1}{2} 
(\eta^{AB}_{a}e^{a}_{i}+\bar{\eta}^{AB}_{a}\bar{e}^{a}_{i})
=\frac{1}{2} \epsilon^{ABCD}E_{C0}E_{Di}. 
\label{condition}
\end{equation}
The required condition is one of those which we impose on 
the anti-symmetric fields $B^{AB}_{\mu\nu}$ of the 2-from 
gravity so that we may introduce vierbeins and derive Einstein 
gravity from it~\cite{2-form gravity}.  
The reason why such conditions arise may be understood by 
the Meissner effect due to the condensation of string fields 
to which $B^{AB}_{\mu\nu}$ couple as gauge 
fields~\cite{condensation of string}. 

We have to clarify this point more explicitly.  In our present case, the
condition (\ref{condition}) gives 
\beq
\frac12 \epsilon^{ABCD} B^{AB}_{i} B^{CD}_{j}=0,
\eeq
which requires that the two metrics in the original 3d gravity are not
independent, but should be identified from the beginning, namely
\beq
g_{ij}=\bar{g}_{ij}, ~~\mbox{where}~~
g_{ij}=\sum_{c} e^{c}_{i}e^{c}_{j}, ~~\mbox{and}~~ \bar{g}_{ij}=\sum_{c}
\bar{e}^{c}_{i}\bar{e}^{c}_{j}.
\label{eqaul metrics}
\eeq
This is a natural constraint to be imposed in the 2N times duplication of
the 3d Einstein gravities. Due to this constraint (20), the number of
degrees of freedom of $B^{AB}_{i}$ is reduced to 12 and can be expressed in
terms of 16 $E^{A}_{\mu}$'s.  

Now the dominant action at low energy induced by the fermion 
pair condensates can be written as 
\begin{equation}
\Delta S_{G} = -f_{g}^{2} \int d^{3}x ~a\sum_{n=1}^{N} 
\frac{1}{4}\epsilon^{ijk}\epsilon_{ABCD}E^{A}_{i}(x, n)
R^{CD}_{k0}E^{B}_{j}(x, n+1) +\cdots.
\end{equation}
Using the expression (\ref{condition}) in the original action
(\ref{original 3D gravity action}), we are lead to
\begin{eqnarray}
& &S_{G}+\Delta S_{G} \nonumber \\
&=&\frac{1}{2a(\kappa_{3})^{2}} \int d^{3}x~a\sum_{n=1}^{N}
\left(\frac14\epsilon^{ijk}\epsilon_{ABCD}E^{A}_{0}(x, n)
R^{CD}_{jk}E^{B}_{i}(x, n) \right. \nonumber \\ 
&+& \left. \{2a(\kappa_{3})^{2}(f_{g})^{2}\} \frac14
\epsilon^{ijk}\epsilon_{ABCD} E^{A}_{i}(x, n) R^{CD}_{k0}E^{B}_{j}(x,
n+1)+\cdots \right). 
\end{eqnarray}
If the following conditions are satisfied,
\begin{equation}
a=
\frac{1}{2(\kappa_{3})^{2}(f_{g})^{2}}~\mbox{and}~(\kappa_{4})^{2}
=2a(\kappa_{3})^{2},
\end{equation}
then we arrive finally at 
\begin{equation}
S_{G}+\Delta S_{G} = 
\frac{1}{2(\kappa_{4})^{2}} \int d^{3}x~a\sum_{n=1}^{N}
E(x, n) R(x, n)+\mbox{(higher order terms in a)},
\end{equation}
where we have the relation between 4d gravity coupling and 4d 
Newton constant as $(\kappa_{4})^{2} = 8 \pi G_{4}$.  
In the gravity case, generation of the extra dimension may 
occur both space like and time like dimensions, since both signs 
are permitted for the generated action $\Delta S_{G}$. 
\section{
Introduction of matter fermions into the gauge model and 
the generation of 4d standard model from 3d one
}
In this section we introduce matter fermions into the previous 
model, especially that of generating 4d gauge theory from 3d ones 
at high energy. 
We will understand that the 4d standard model is promising 
to be generated in this way.  
In the following we restrict ourselves to the less challenging 
version of generating one "space-like" dimension of $x^{3}$. 
Accordingly, we will use $\{i, j, k, \cdots\}$ for 3d indices 
$\{0,1,2\}$, and $\{\mu, \nu, \lambda, \cdots\}$ for 4d ones.

Then, our starting action of matter fermions at high energy is 
\begin{equation}
S_{f} =\int d^3x \sum_{n=1}^{N} \Bigl(2aK\bar{\psi}(x, n)
i\gamma^{i}D_{i}(A(n))\psi(x, n) 
-\bar{\psi}(x, n)\psi(x, n) \Bigr),
\end{equation}
where we have introduced the hopping parameter $K$ in place of 
the mass parameter, following Wilson~\cite{lattice theory}.  
In 3d, fermions are 2 component, so that we have to replace 
$\psi$ by a doublet of the chiral components 
$\psi_{L}$ and $\psi_{R}$, namely $\psi=(\psi_{L}, \psi_{R})^{T}$.  
If the chiral partner does not exist, then we consider the partner 
to be $0$.  
Even without having $\gamma_{5}$ in 3d, 2 component fermions labelled 
by $L$ and $R$ are eigenstates of the helicity operator 
$h=\bsigma\cdot {\bf p}/|{\bf p}|$. 

More explicitly, the action reads
\begin{eqnarray}
S_f = \int d^3x \sum_{n=1}^{N} \Bigl(
& 2aK{\psi}_{L}(n)^{\dagger} i(D_{0}-
\bsigma{\bf D})\psi_{L}(n)& 
\nonumber \\
&+2aK{\psi}_{R}(n)^{\dagger} i(D_{0}+\bsigma{\bf D})\psi_{R}(n)& 
\nonumber \\
&-{\psi}_{R}(n)^{\dagger}\psi_{L}(n) - 
\psi_{L}(n)^{\dagger}\psi_{R} (n)
&\Bigr).
\end{eqnarray}
The action is simply N times multiplication of the 3d fermionic 
gauge invariant actions of the group $(G)^{N}$. 

After the condensation occurs at the energy lower than $\Lambda_{s}$,  
the extra gauge field of $A_{3}(x; n)$ can be generated as
\begin{equation}
U(x; n, n+1)=e^{iaA_{3}(x, n)}.
\end{equation}
Then, the additional actions $\Delta_{1}S_{f}$ and 
$\Delta_{2}S_{f}$ may appear:
\begin{eqnarray}
\Delta_{1}S_{f} &=& \int d^3x \sum_{n=1}^{N} K 
\Bigl(\bar{\psi}(n)i\gamma^{3}U(n, n+1)\psi(n+1) 
\nonumber \\ 
&&~~~~
-\bar{\psi}(n+1)i\gamma^{3}U(n, n+1)^{\dagger} \psi(n) 
\Bigr) \nonumber \\
&=& \int d^3x \sum_{n=1}^{N} K\Bigl(\psi_{L}(n)^{\dagger}
i(-\sigma^{3})U(n, n+1)\psi_{L}(n+1) 
+\psi_{R}(n)^{\dagger} 
i\sigma^{3}U(n,n+1)\psi_{R}(n+1) \nonumber \\
&& \hspace{-1cm}
-\psi_{L}(n+1)^{\dagger} i(-\sigma^{3})U(n,n+1)^{\dagger} 
\psi_{L}(n)-\psi_{R}(n+1)^{\dagger}i\sigma^{3} U(n;n+1)^{\dagger} 
\psi_{R}(n)\Bigr),
\end{eqnarray}
and
\begin{eqnarray}
\Delta_{2}S_{f} &=& \int d^3x \sum_{n=1}^{N} K 
\Bigl( \bar{\psi}(n)U(n, n+1)\psi(n+1)
+\bar{\psi}(n+1) U(n, n+1)^{\dagger} \psi(n) \Bigr) 
\nonumber \\
&=& \int d^3x \sum_{n=1}^{N} K\Bigl(\psi_{R}(n)^{\dagger}
U(n,n+1)\psi_{L}(n+1)+\psi_{L}(n)^{\dagger}U(n, n+1)\psi_{R}(n+1) 
\nonumber \\
& &+\psi_{L}(n+1)^{\dagger} U(n,n+1)^{\dagger}\psi_{R}(n)
+\psi_{R}(n+1)^{\dagger}U(n,n+1)^{\dagger}\psi_{L}(n)\Bigr).
\end{eqnarray}
Now, we can understand that the first additional action 
$\Delta_{1}S_{f}$ leads to the fermion's kinetic term as well as 
the gauge interaction with $A_{3}$, namely
\begin{eqnarray}
\Delta_{1}S_{f} &=& 
\int d^3x \sum_{n=1}^{N} \Bigl(2aK 
\left(
\bar{\psi}(n)i\gamma^{3}\frac{1}{2a} [\psi(n+1)-\psi(n-1)]\right)
\nonumber \\
&&
-aK\left(\bar{\psi}(n)\gamma^{3}A_{3}\psi(n+1)+\bar{\psi}(n+1)
\gamma^{3}A_{3}\psi(n) \right) \Bigr) +\cdots.
\end{eqnarray}
The second additional action $\Delta_{2}S_{f}$ is that Wilson 
proposed a long time ago.  
This action is combined with the mass-like term in the original 
action, giving
\begin{equation}
\Delta_{2}S_{f}=\int d^3x \sum_{n=1}^{N}
\left(K\left(\bar{\psi}(n)\psi(n+1)+\bar{\psi}(n+1)\psi(n)\right)
-\bar{\psi}(n)\psi(n)\right).
\end{equation}
Combining these generated actions with the original fermion action, 
we obtain the 4d gauge invariant fermion action having higher 
order corrections in $a$. 
The propagator of the fermion $S_{F}$ reads 
\begin{equation}
S_{F}({\bf p},
p_{3})=\frac{i}{\sum_{i=0}^{2}\gamma^{i}p_{i}+\gamma^{3}
\frac{\sin(p_{3}a)}{a}-m},
\end{equation}
having the mass
\begin{equation}
m=\frac{1-2K\cos(p_{3}a)}{2aK}.
\end{equation}
If the theory is left-right symmetric, the above reasoning works 
well, and we have no doubling problem as is shown by 
Wilson~\cite{lattice theory}: 
The particle with momentum $p_{3} \sim 0$ has mass $(1-2K)/2aK$, 
while the doubling partner (having opposite chirality) with 
$p_{3} \sim \pi/a$ has the larger mass of  $(1+2K)/2aK$, so that 
for sufficiently small $a$, the unwanted partner can be decoupled.  
Therefore, the left-right symmetric 4d QED and QCD can be reproduced 
without any troubles.

Next, we try to generate the 4d standard model from 3d ones. 
The standard mode is a gauge theory based on the group 
$SU(3)_C \times SU(2)_L \times U(1)_Y$, in which the fermions 
are left-right asymmetric. 
Even though the first additional action $\Delta_{1}S_{f}$ can be 
generated, by reproducing correctly the 4d standard model "gauge 
interactions" with different link variables $U_{L}(x; n, n+1)$ and
$U_{R}(x; n, n+1)$ for L- and R-handed sectors, the second additional
action $\Delta_{2}S_{f}$ can 
not be generated properly in the left-right asymmetric part of 
$SU(2)_L \times U(1)_Y$.

In order to remedy this point, we introduce the other fermion pair 
condensates $H_{0}(x; n, n)$ and $H_{1}(x; n, n+1)$ in addition to
$U_{L}(x; n, n+1)$ and $U_{R}(x; n, n+1)$. 
Namely, we replace the gauge fields by the Higgs fields, and 
supply the mass term $\Delta_{0}S_{f}$ and the Wilsonian action 
$\Delta_{3}S_{f}$ in the left-right 
asymmetric theory.

In the standard model, we have these actions as
\bea
\Delta_{0}S_{f} &=& -\int d^{3}x \sum_{n=1}^{N} a 
\biggl( q_{L}(n)^{\dagger} H_{0}(n,n) u_{R}(n) 
\nonumber \\
&& ~~~~~~~~~~~~
+q_{L}(n)^{\dagger} \tilde{H}_{0}(n, n) d_{R}(n)
+ h.c. \biggr), 
\\
\Delta_{3}S_{f} &=& \int d^{3}x \sum_{n=1}^{N} a 
\biggl( q_{L}(n)^{\dagger} H_{1}(n,n+1) u_{R}(n+1) 
\nonumber \\
&& ~~~~~~~~~~~~
+q_{L}(n)^{\dagger} \tilde{H}_{1}(n, n+1) d_{R}(n+1)
+ h.c. \biggr), 
\eea
where $H_{i}=(\phi^{0}_{i}, \phi^{-}_{i})^{T}~~(i=0, 1)$ is the two types
of Higgs 
doublets and as usual  $\tilde{H}_{i}=(-\phi^{+}_{i},
\phi_{i}^{0}*)^{T}~~(i=0, 1)$.

Then, the vacuum expectation values of the Higgs, 
$\langle \phi^{0}_{i} \rangle = v_{i}~~(i=1, 2)$, 
give the mass of $u$ and $d$ quarks as
\begin{equation}
m_{u}=\frac{v_{0}-v_{1}\cos(p_{3}a)}{K_{u}},
~~\mbox{and}~~m_{d}=\frac{v_{0}-v_{1}\cos(p_{3}a)}{K_{d}},
\end{equation}
and the Yukawa couplings as
\begin{equation}
y_{u}=\frac{v_{0}-v_{1}\cos(p_{3}a)}{(v_{0}-v_{1})K_{u}},~~\mbox{and}~~y_{d}=
\frac{v_{0}-v_{1}\cos(p_{3}a)}{(v_{0}-v_{1})K_{d}}.
\end{equation}
The doubling particles with $p_{3} \sim 0$ and $p_{3} \sim \pi/a$ 
has the opposite sign in the $v_{1}$ terms, but not in the $v_{0}$ terms. 

Therefore, if the following inequality holds
\beq
0 \le v_{0}-v_{1} \ll v_{0}+v_{1},
\label{inequality}
\eeq
the doubling problem between particles with 
$p_{3} \sim 0$ and $p_{3}\sim \pi/a$ can be solved even in 
our simple usage of the Wilson fermions. However, there still remains the
problem of whether we may find a proper condensation mehcanism supporting
this inequality (\ref{inequality}). 

It is important to recognize here that we have introduced two sets of
discrete lattices for L- and R-handed fields separately, and that the two
types of Higgs fields are assumed to be generated.  If we consider the
direction of separating L-handed lattice points from the R-handed ones to
be the 5th dimension, out treatment of fermions may be related to the
recent developments such as the domain wall 
fermion~\cite{domain wall fermion} and so on. 

Anyway, in the promising scenario of deriving 4d standard model 
at low energy from 3d ones at high energy the Higgs field 
(fundamental representation) as well as an extra component of 
the gauge fields (adjoint representation) should be generated at 
low energy as the fermion pair condensates due to the strong 
interactions based on the group $G_{s}$. 

Although we have not specify the strong interaction sector 
explicitly, it may be a kind of the "chiral model".  
It is because we know well that the scalar condensate of fermions 
is produced as the $\sigma$ meson which corresponds to $H$ (Higgs 
scalar) in our case, while the pseudo scalar condensates of 
$\pi$ mesons correspond to our $U$ (an extra component of 
gauge fields).  

Finally we will comment on the fact that the the gauge field 
$U(x; n, n+1)$ is the connection of the discrete lattice on 
which the same chirality fields are located.  
On the other hand, the Higgs fields $H_{0}(x; n, n)$ and $H_{1}(x; n, n+1)$
can be 
considered also the ``connections" of the discrete lattices on 
which the opposite chirarity fields are located.  
This remind us of the treatment by Connes of the Higgs 
scalar~\cite{noncommutative geometry} in which the Higgs scalar is the
connection of the discrete lattice in the non-commutative 
geometry. 
Matter coupling of fermions when the gravity is switched on 
may be studied similarly. For example, we may start from the following 3d
fermionic action coupled with gravity:
\beq
S_{G}^{f}=\int d^{3}x~~\sum_{n=1}^{N} (detE^{A}_{\mu}(x; n) )
aK\left(\bar{\psi}(x; n) i\gamma^{A} \sum_{i=0}^{2} E^{i}_{A}(x; n)
D_{i}(\omega(x; n)) \psi(x; n) + \mbox{h.c} \right ),
\eeq
where the covariant derivative is given by
\beq
D_{i}(\omega(x; n)) = \partial_{i} +\frac{1}{8} [ \gamma^{A}, \gamma^{B}]
\omega^{AB}_{i}(x; n).
\eeq
If the link variable $O(x; n, n+1)$, connecting $n$ and $n+1$ lattice
points, is generated as 
\beq
 O(x; n, n+1)= e^{\frac{a}{8} [ \gamma^{A}, \gamma^{B}]
\omega^{AB}_{\bar{3}}(x; n)},
\eeq
Then the following action may be induced:
 \bea
\Delta S_{G}^{f}&=&\int d^{3}x~~\sum_{n=1}^{N} (detE^{A}_{\mu}(x; n) ) 
\nonumber \\
&& \times K\left(\bar{\psi}(x; n) i\gamma^{A}E^{3}_{A}(x;
n)E^{\bar{3}}_{3}(x; n) O(x; n, n+1) \psi(x; n) + h.c. \right ).
\label{fermion with gravity}
\eea
The induced action supplies the missing contribution from the 3rd axis:
\bea
\Delta S_{G}^{f}&=&\int d^{3}x~~\sum_{n=1}^{N} (detE^{A}_{\mu}(x; n) ) 
\nonumber \\
&& \Bigl( 2aK\left(\bar{\psi}(x; n) i\gamma^{A}E^{3}_{A}(x;
n)E^{\bar{3}}_{3}(x; n) \frac{1}{2a}[\psi(x; n+1)- \psi(x; n-1)] \right ) 
\nonumber \\
&& +aK\left(\bar{\psi}(x; n) i\gamma^{A}E^{3}_{A}(x; n)E^{\bar{3}}_{3}(x;
n) \frac{1}{8}[ \gamma^{A}, \gamma^{B}] \omega^{AB}_{\bar{3}}(x; n) \psi(x;
n+1) + \mbox{h.c.}\right) \Bigr) 
\nonumber \\
&&+ \cdots.
\eea
In the above expressions, $E^{\bar{3}}_{3}$ is necessary for converting the
flat space index $\bar{3}$ to the curved space one $3$. 


\section{Conclusion}
In this paper, by generalizing the work by Arkani-Hamed et 
al.~\cite{Georgi}, generation of 4d gauge theories and gravity 
at low energy from the 3d ones at high energy is studied. 
In the gauge theories, the generation of the space-like dimension 
is found to be easier than that of the time-like one.  
Referring to the  2-form gravity (or the Ashtekar formalism)~\cite{2-form
gravity}, \cite{condensation of string}, the 4d Einstein gravity 
is shown to be generated at low energy from the $2N$ times 
multiply of the 3d Einstein gravity.  In the derivation we have to impose a
constraint that the two metrics prepared in the biginning should be
identical.  For the gravity there is no difference between the generation 
of time-like and space-like dimensions.  
The fact that the 4d gravity becomes 3d one at high energy is 
thought to be something to do with the holographic 
principle~\cite{holographic principle}.

Introduction of matter fermions is studied following Wilson with 
the so called Wilson fermions~\cite{lattice theory}.  
The matter couplings are correctly reproduced when coming down 
from 3d theory to 4d one.  
To reproduce the standard model interactions two kinds of fermion 
condensates should be produced. 
One of them becomes the extra component of the gauge fields and 
the other becomes the Higgs fields at low energy.  
In this way the left-right asymmetric interactions such as 
the 4d standard model interactions are possible to be generated 
at low energy, starting from the multiplicated 3d ones at high 
energy.  
The doubling problem of having the opposite chiral 
partner may be solved in the left-right asymmetric part (weak 
interactions), as well as in the left-right symmetric parts 
(electromagnetic and strong interactions).  
For this purpose we have prepared two kinds of discrete lattices,
separately for L- and R-handed fields, and assumed the generation of two
types of Higgs fields.  
In order to give the firm foundation to our models, it should 
be clarified the dynamics of how the fermion pair condensation 
occurs so as for the gauge fields, the Higgs fields and the spin 
connections of gravity to be properly produced. 

The most interesting next problem is how these theories are seen 
in high energy experiments and observations. Probably the scale 
of the metric's (gravity) generation $\Lambda_{s}'$ is 
sufficiently larger than that of the standard model's generation 
scale, $\Lambda_{s}$.  
The Feynman rules of gauge theories and gravity become modified 
when the lattice constant will manifest itself at high energy. 
Using these modified rules, we have to clarify the constraints 
from the available high energy experiments and observations.  
At the same time we have to show the experimental and 
observational prospects of our models in the future.

%
%
\section*{Acknowledgment}
The author is grateful to  Gi-Chol Cho who introduces this new 
topics to the author by suggesting the possibility of 
generating 4d theory from 3d one, having useful discussions and 
reading the manuscript.  
He also thanks Etsuko Izumi for her collaboration with him. 

After this paper was submitted for publication, the author was informed of
the various related works previously performed~\cite{related works}. He
gives his thanks to the individual authors for this information.
%
%

\end{document}